\documentclass[aps,prc,twocolumn,showpacs,floatfix,nofootinbib,preprintnumbers,superscriptaddress,amsmath,amssymb,longbibliography,amsproc]{revtex4-1}
\usepackage{epsfig,dsfont,amssymb,amsmath,amsthm,amsfonts,amsbsy,mathrsfs}
\usepackage{graphicx}
\usepackage{amsmath}
\usepackage{amssymb}%
\usepackage{dcolumn}
\usepackage{multirow}
\usepackage[colorlinks=true,linkcolor=blue,citecolor=blue,urlcolor=blue]{hyperref}
\usepackage{bbold}
\usepackage{bibentry}

\graphicspath{{./}}

\usepackage[disable]{todonotes}

\newcommand{\be}{\begin{equation}}
\newcommand{\ee}{\end{equation}}
\newcommand{\ba}{\begin{eqnarray}}
\newcommand{\ea}{\end{eqnarray}}

\begin{document}

\title{Entanglement entropy of nuclear systems}

\author{Chenyi Gu}
\affiliation{Department of Physics and Astronomy, University of
  Tennessee, Knoxville, Tennessee 37996, USA}

\author{Z.~H.~Sun}
\affiliation{Physics Division, Oak Ridge National Laboratory, Oak
  Ridge, Tennessee 37831, USA}

\author{G.~Hagen}
\affiliation{Physics Division, Oak Ridge National Laboratory, Oak
  Ridge, Tennessee 37831, USA}

\affiliation{Department of Physics and Astronomy, University of
  Tennessee, Knoxville, Tennessee 37996, USA}

\author{T.~Papenbrock}
\affiliation{Department of Physics and Astronomy, University of
  Tennessee, Knoxville, Tennessee 37996, USA}

\affiliation{Physics Division, Oak Ridge National Laboratory, Oak
  Ridge, Tennessee 37831, USA}

\begin{abstract}
We study entanglement entropies between the single-particle states of the hole space and its complement in nuclear systems. Analytical results based on the coupled-cluster method show that entanglement entropies are proportional to the particle number fluctuation and the depletion number of the hole space for sufficiently weak interactions. General arguments also suggest that the entanglement entropy in nuclear systems fulfills a volume instead of an area law. We test and confirm these results by computing entanglement entropies of the pairing model and neutron matter, and the depletion number of finite nuclei. 
\end{abstract}
\maketitle

\section{Introduction}
\label{sec:intro}

Entanglement is a key property in quantum mechanics~\cite{horodecki2009}. It refers to non-local aspects of a wave function and usually makes it hard to numerically solve a quantum many-body problem.
Expressions such as ``wave-function correlations'' or ``fluctuations" are often used as synonyms for entanglement. However, the latter has the advantage that it can be quantified using entropies. In this article, we are interested in entanglement entropies of ground states in neutron matter and nuclear models that arise when the single-particle basis is partitioned into two complementary sets.      

Entanglement is widely studied in different areas of physics~\cite{eisert2010}. In shell-model calculations, understanding entanglement helps when applying the density-matrix renormalization group~\cite{legeza2015,Tichai2022}. Recently, advances in quantum information science and quantum computing also renewed an interest in exploring entanglement in nuclear systems~\cite{beane2019, robin2021, faba2021, kruppa2022,Pazy2022,bai2022,lacroix2022,bulgac2022,johnson2022}. A better understanding of entanglement might thus benefit both classical and quantum computations of atomic nuclei.

Let us define those metrics that quantify the entanglement of quantum systems. We assume that the Hilbert space ${\cal H}$ is decomposed as a  ${\cal H} = {\cal H}_A \otimes {\cal H}_B$ in terms of the Hilbert spaces of two subsystems $A$ and $B$.
The density matrix of the ground state $|\Phi\rangle$ is 
\be
\rho = |\Phi\rangle\langle\Phi| \ , 
\ee
and the reduced density matrix of the subsystem $A$ is obtained by tracing over the subsystem $B$, i.e. 
\be
\rho_{A} = \operatorname{Tr}_B \rho \ .
\ee
The density matrices $\rho_A$ and $\rho$ are Hermitian,  non-negative (i.e. they have non-negative eigenvalues), and fulfill $\operatorname{Tr}\rho=1$. And we say $\rho_A$ is entangled with $B$ when it can not be represented by a pure state, i.e.,   $\operatorname{Tr}\rho_A^2<1$. Measures such as entropy or  mutual information can be used to quantify the entanglement. In this paper, we consider the R\'enyi entropy~\cite{renyi1961} 
\begin{equation}
\label{eq:renyi}
    S_{\alpha}=\frac{1}{1-\alpha} \ln \operatorname{Tr} \rho^{\alpha}_A \ .
\end{equation}
Here 
$\alpha \in (0,1)\cup (1,\infty)$, 
and the von~Neumann entropy arises as the limiting case of the R\'enyi entropy for $\alpha\to 1$,  i.e.  
\begin{equation}
\label{SvN}
    S_1=\underset{\alpha\to 1}{\lim} S_\alpha = -\operatorname{Tr}(\rho_A \ln \rho_A) \ .
\end{equation}

\begin{figure}[!htbp]
\includegraphics[width=0.4\textwidth]{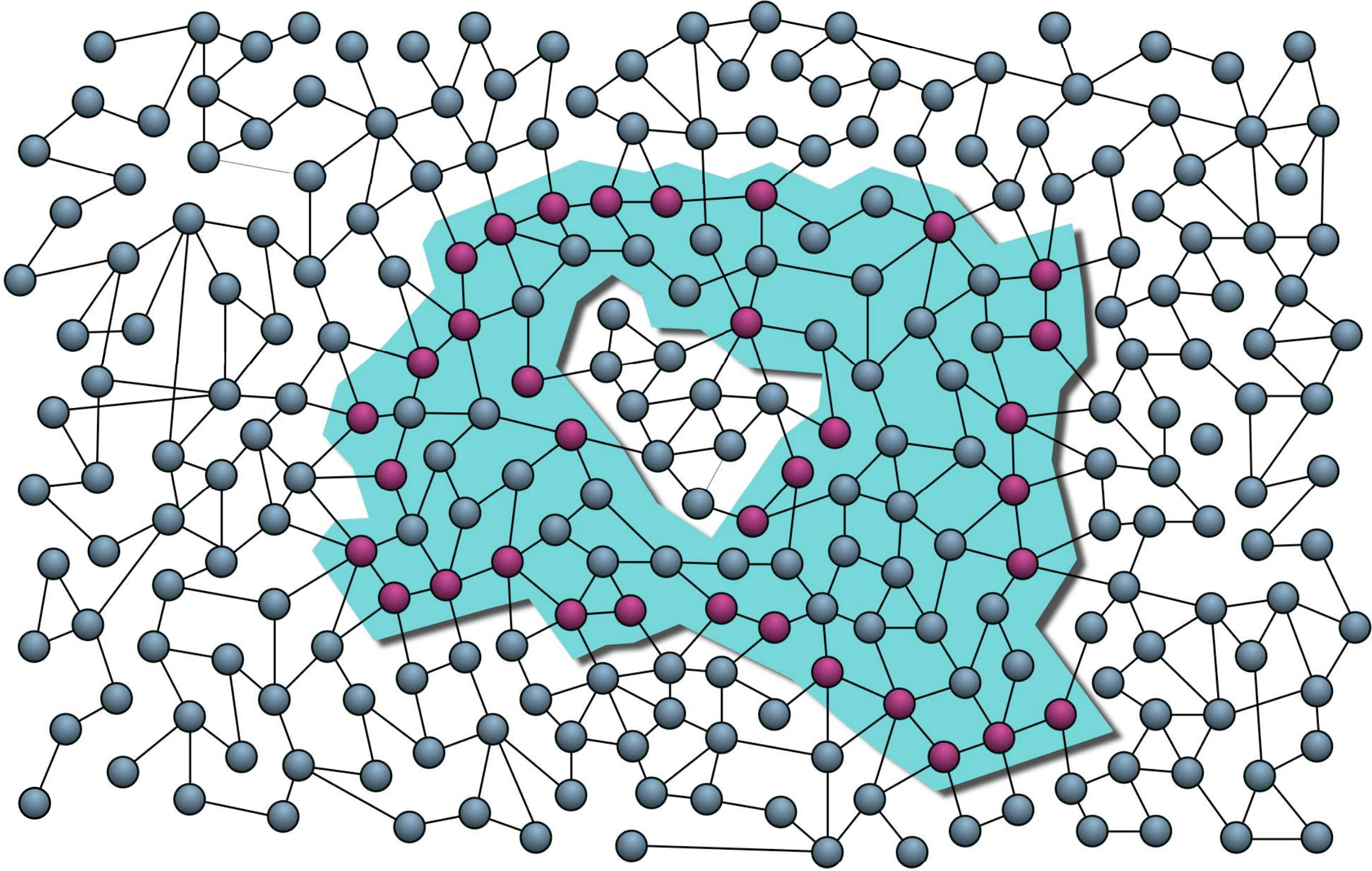}
\caption{Lattice system (sites and links) partitioned into two regions (colored blue and white). The red sites in the blue region have links to sites in the white region. Taken from Ref.~\cite{eisert2008} with permission of the authors; see also Ref.~\cite{eisert2010}.}
\label{fig:eisert}
\end{figure}

In lattice systems with local interactions, one often finds that the entanglement entropy grows proportional with the area (times some logarithmic corrections) when the system is partitioned into two subsystems~\cite{eisert2010}. Figure~\ref{fig:eisert} shows how this meets expectations. The red-colored sites within the blue subsystem have links to the white complement, and their number is proportional to the size of the boundary. This leads to an area law for entanglement entropy in three dimensions. 

\textcite{wolf2006} and \textcite{gioev2006} showed that the von Neumann entanglement entropy for fermionic tight-binding Hamiltonians and free fermions in $d$ dimensions, respectively, scales as $S_1\sim L^{d-1}\log{L}$, where $L$ is a linear dimension of subsystem $A$. Thus, these fermionic systems fulfill area laws with logarithmic factors. 
\textcite{gioev2006} and \textcite{klich2006} also showed that the particle-number variation  $(\Delta N)^2$ gives upper and lower bounds of the von~Neumann entropy via
\begin{equation}
    4(\Delta N)^2 \leq S_1 \leq \mathcal{O}(\log L)(\Delta N)^2 \, .
\end{equation}
\textcite{leschke2014} extended the proof to general  R\'enyi entanglement entropies $S_\alpha$. 
Extensions to interacting (and exactly solvable systems) can be found in Refs.~\cite{barthel2006, plenio2005}.  \textcite{masanes2009} pointed out that area laws with logarithmic factors hold for a fermionic state if ``(i) the state has sufficient decay of correlations and (ii) the number of eigenstates with vanishing energy density is not exponential in the volume.'' 

\begin{figure}[!htbp]
\includegraphics[width=0.4\textwidth]{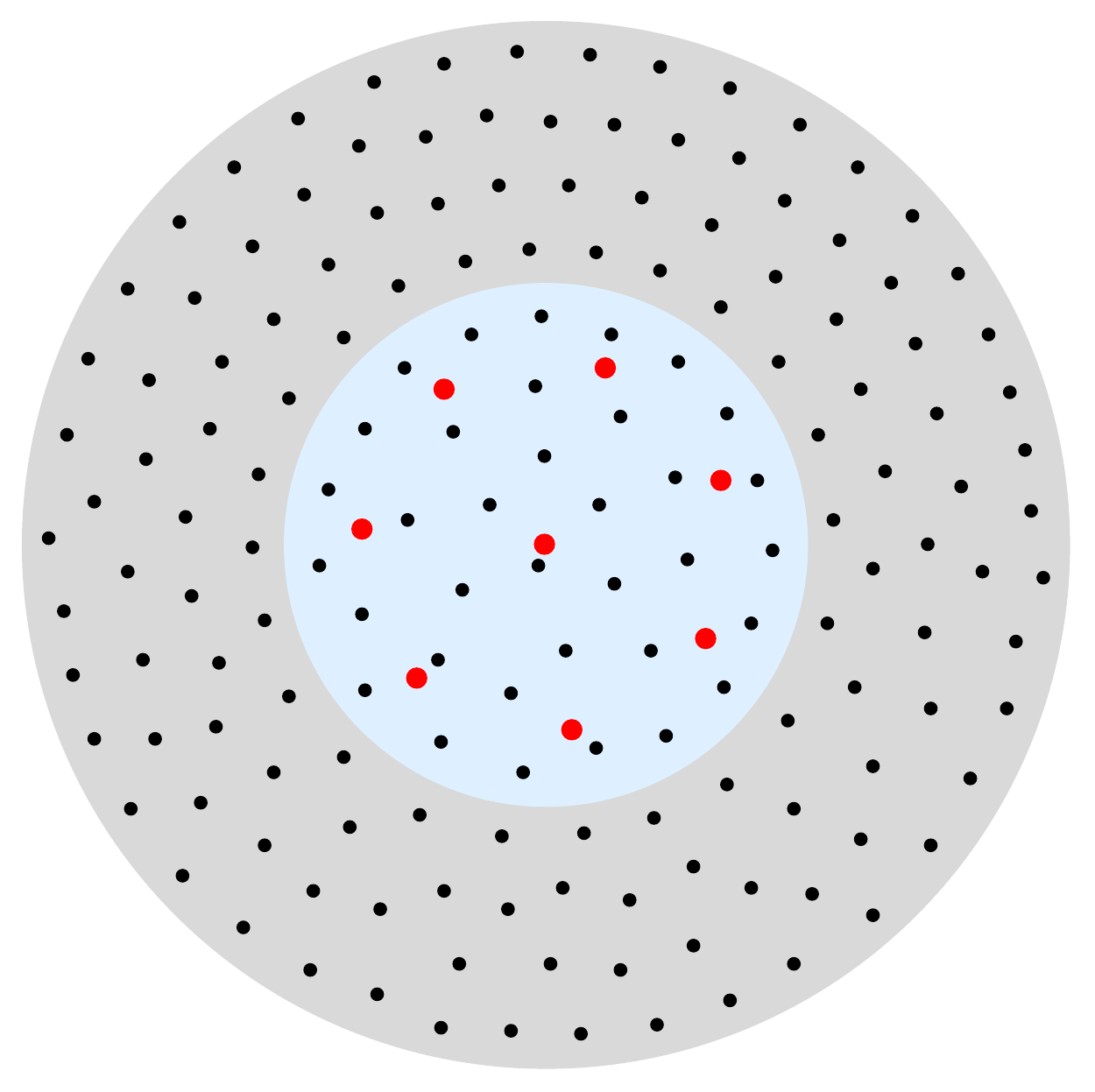}
\caption{Position-space sketch of the nuclear volume (depicted in light blue) and its complement (depicted in gray) for a finite spherical basis. The red points represent (localized) hole states while the black points symbolize localized particle states. The former (latter) exhibits a nearest neighbor distance that is inversely proportional to the Fermi momentum (momentum cutoff). Thus, one expects a volume law for the entanglement entropy between particle and hole states.}
\label{fig:localized}
\end{figure}

While the first condition is expected to be fulfilled for atomic nuclei, the second seems not. After all, nuclei are open quantum systems and resonant and scattering states are abundant. A question also arises about how to partition the Hilbert space when dealing with a finite system. 
We partition the system into the single-particle states of the reference state (the hole space) and its complement (the particle space). This partition results, e.g., from a Hartree-Fock computation or from a naive filling of the spherical shell model. The single particle states in both subspaces are usually delocalized in position space. Hartree-Fock orbitals, for instance, are localized on an energy surface in phase space but spread out in position space. One can now imagine using unitary basis transformations in the hole and particle spaces such that single-particle states become localized in both partitions~\cite{foster1960,edmiston1963,hoyvik2012}. (Orthogonality requirements might lead to somewhat less localized single-particle states, though.) The ideal situation is depicted in Fig.~\ref{fig:localized}.
Here, the red points are the hole states in position space. Their nearest neighbor distance is about $\pi/k_F$ where $k_F$ is the Fermi momentum. The ``volume'' occupied by the reference state is depicted in light blue. The region outside the nuclear volume is depicted in light gray. The black points denote the states of the particle space. Their nearest-neighbor distance is about $\pi/\Lambda$ where $\Lambda$ denotes the momentum cutoff. Thus, their density in position space is larger than the density of the red hole states and the resolution of the finite-Hilbert-space identity also demands that there is a considerable number of particle states ``inside'' the volume occupied by the nucleus. (The density of localized states in the grey and light blue areas is equal.) Even for a short-ranged (and possibly local) nuclear interaction, we see that every hole state is correlated with particle states. Thus, we expect a volume law for the entanglement entropy between particle and hole space. 

This expectation also holds in momentum space. There, the hole states occupy the Fermi sphere (evenly distributed) while the particle states occupy the complement. As the nuclear interaction is short-ranged in position space, it becomes long-ranged in momentum space and thereby also leads to a volume law for entanglement entropy.

Similar expectations also hold for lattice computations of atomic nuclei~\cite{lee2009} where the single-particle basis consists of a cubic lattice in position space. Let us consider a nucleus with an average density $n_0\approx 0.16$~fm$^{-3}$. The nucleus with mass number $A$ occupies a volume $A/n_0$ and the number of available single-particle states inside this volume 
\be
\Omega = g_\mathrm{st}\frac{A}{a^3 n_0} \ ,
\ee
where $a$ is the lattice spacing and $g_{\rm st}=4$ the spin/isospin degeneracy. The reference state of the nucleus consists of $A$ single-particle states (also occupying the volume $A/n_0$).  We have
\be
\label{nminA}
\Omega-A = \Omega\left({g_\mathrm{st}\over a^3 n_0}-1\right)
\ee
and for typical lattice spacing $a=1.3$~fm or $a=2$~fm~\cite{elhatisari2016, lu2019}, we find $\Omega-A\approx 10A$ and $2A$, respectively. Thus we expect a volume law for the entanglement entropy. 

We also note that $\Omega \sim a^{-3}$ for $a\to 0$ and recall that the ultraviolet cutoff is $\Lambda=\pi/a$. Thus, entanglement is expected to increase with increasing cutoff of the nuclear interaction.

This paper is organized as follows. In Sec.~\ref{Sec:analytical}, we present analytical results for the entanglement entropy in finite systems. In Sec.~\ref{sec:nuclear} we test our predictions and present results for the pairing model,  neutron matter, and finite nuclei.

\section{Analytical results}
\label{Sec:analytical}
In this Section, we utilize coupled-cluster theory~\cite{kuemmel1978,bishop1991,bartlett2007,hagen2014} to derive analytical results for the R{\'e}nyi entropy,  the particle fluctuation of the hole space, and their mutual relation. 

\subsection{Coupled-cluster theory}
Following the standard coupled-cluster formulations, for a many-body system with $N$ fermions, we express the ground state wavefunction $|\Psi\rangle$ as 
\begin{equation}
\label{eq:CCground}
    |\Psi\rangle = e^{\hat{T}}|\Phi\rangle \;,
\end{equation}
using the reference state 
\begin{equation}
    |\Phi\rangle = \prod_{i=1}^N \hat{a}^\dagger_i |0\rangle \ .
\end{equation}
The cluster operator $\hat{T}=\hat{T}_1+\hat{T}_2+\cdots+\hat{T}_N$ contains all possible $k$-particle--$k$-hole excitations 
\begin{equation}
    \hat{T}_{k}=\frac{1}{(k !)^{2}} \sum_{i_{1}, \ldots, i_{k} ; \atop a_{1}, \ldots, a_{k}} t_{i_{1} \ldots i_{k}}^{a_{1} \ldots a_{k}} \hat{a}_{a_{1}}^{\dagger} \ldots \hat{a}_{a_{k}}^{\dagger} \hat{a}_{i_{k}} \ldots \hat{a}_{i_{1}} \;.
\end{equation} 
Here the indices $i_k$ and $a_k$ represent occupied (hole) and unoccupied (particle) orbitals respectively.
We use the convention that indices $i, j$ and $a,b$ refer to hole and particle states, respectively. To obtain the coupled-cluster amplitudes $t_{i_{1} \ldots i_{k}}^{a_{1} \ldots a_{k}}$, we solve the  amplitude equations
\begin{equation}
\label{CCamps}
    \left\langle\Phi_{i_1i_2 \ldots}^{a_1a_2 \ldots}|e^{-\hat{T}}\hat{H}e^{\hat{T}}|\Phi_0\right\rangle = 0
\end{equation}
where
\begin{equation}
    \left|\Phi_{i_1i_2 \ldots}^{a_1a_2 \ldots}\right\rangle  \equiv\hat{a}_{a_1}^{\dagger} \hat{a}_{a_2}^{\dagger}\cdots \hat{a}_{i_2} \hat{a}_{i_1} \left|\Phi_{0}\right\rangle
\end{equation}
and then compute the energy via 
\begin{equation}
\label{Eccd}
    E = \left\langle\Phi|e^{-\hat{T}}\hat{H}e^{\hat{T}}|\Phi\right\rangle  \ .
\end{equation}

For the purpose of analyzing results of the pairing model and neutron matter, we use the coupled cluster doubles (CCD) approximation. Here the cluster operator is $\hat{T}=\hat{T_2}$, and the ground state becomes
\begin{equation}
\label{eq:CCD_approximation}
    |\Psi_{\rm CCD}\rangle = \exp(T_2)|\Phi\rangle \ . 
\end{equation}
The omission of singles (i.e. 1-particle--1-hole excitations) is valid because the pairing-model Hamiltonian only changes the occupation of pairs and because neutron matter is formulated in momentum space where the conservation of momentum forbids single-particle excitations. For other finite systems, the contributions of singles are small in the Hartree-Fock basis. 
The $N$-body density matrix associated with the ground state is 
\begin{equation}
\label{eq:rho}
    \hat{\rho} = \frac{\left|\Psi_{\mathrm{CCD}}\rangle\langle\Psi_{\mathrm{CCD}}\right|}{\langle\Psi_{\mathrm{CCD}}|\Psi_{\mathrm{CCD}}\rangle} \ .
\end{equation}

Since we separate particles and holes we can express states as the following products, 
\begin{equation}
    \left|\Phi_{i_1i_2 \ldots}^{a_1a_2 \ldots}\right\rangle =  |a_1 a_2 \cdots \rangle \otimes |i_1^{-1} i_2^{-1} \cdots \rangle \ .
\end{equation}
The hole-space reduced density matrix $\rho_{\rm H}$ is obtained by tracing the density matrix $\rho$ over the particle states. 
The  matrix elements of $\rho_{\rm H}$ are
\begin{align}
\label{rhoH}
 \langle |\rho_{\rm H}|\rangle &=  \langle\Phi|\hat{\rho}|\Phi\rangle\ , \nonumber\\
 \langle i_1^{-1} i_2^{-1}| \rho_{\rm H}|j_1^{-1} j_2^{-1}\rangle &= 
 \sum_{a_1<a_2} \langle \Phi_{i_1i_2}^{a_1a_2}|\hat{\rho}|\Phi_{j_1j_2}^{a_1a_2}\rangle \ , \nonumber\\
 &\vdots \  \nonumber\\
\langle i_1^{-1} \cdots i_N^{-1}| \rho_{\rm H}|j_1^{-1} \cdots j_N^{-1}\rangle &= \nonumber\\
\sum_{a_1<\cdots <a_N}&\langle \Phi_{i_1\cdots i_N}^{a_1\cdots a_N}|\hat{\rho}|\Phi_{j_1\cdots j_N}^{a_1\cdots a_N}\rangle \ . 
\end{align}
The first line in Eq.~(\ref{rhoH}) is obtained by tracing over the vacuum state in the particle space, and the second line results from tracing over two-particle states; for the last two lines the trace is over $N$-particle states. As we use the CCD approximation, all traces over odd-numbered particle states vanish. We can easily check that $\operatorname{Tr}\rho_{\rm H}=1$.

\subsection{Approximate entropies}
The exact evaluation of all matrix elements is challenging and we make the approximation
\begin{equation}
\label{CCDex}
\begin{aligned}
    |\Psi_{\mathrm{CCD}}\rangle &\approx \left(1+\hat{T}_2\right)|\Phi\rangle \\
    &=|\Phi\rangle + \frac{1}{4}\sum_{abij} t_{ij}^{ab}|\Phi_{ij}^{ab}\rangle\;.
\end{aligned}
\end{equation}
assuming that $\hat{T}_2$ is small in  a sense we specify below. Thus, we obtain the $\hat{T}_2$  amplitudes from the solution of the coupled-cluster equations but only employ the linearized approximation of the wave function for the computation of the density matrix. Then,  
\begin{equation}
    \hat{\rho} = C^{-1}\left|\Psi_{\mathrm{CCD}}\rangle\langle\Psi_{\mathrm{CCD}}\right| \ ,
\end{equation}
with the normalization coefficient
\begin{eqnarray}
\label{norm}
    C &\equiv& \langle\Psi_{\mathrm{CCD}}|\Psi_{\mathrm{CCD}}\rangle \nonumber\\
    &=& 1+t^2\ .
\end{eqnarray}
Here we used the shorthand
\begin{equation}
\label{eq:t2}
    t^2 \equiv \frac{1}{4}\sum_{i j a b}t_{i j}^{a b}t_{i j}^{a b} \;.
\end{equation}
The approximation~(\ref{CCDex}) is valid for $t^2\ll 1$, and this quantifies in what sense $\hat{T}_2$ is small. 
Tracing over the particle space yields the reduced density matrix
\begin{equation}
    \hat{\rho_{\mathrm{H}}} = {1\over C} \left( |\rangle\langle| + 
    \sum_{a<b} t_{ij}^{ab} t_{kl}^{ab} \left| k^{-1} l^{-1}\right\rangle \left\langle j^{-1} i^{-1} \right|\right) \ .
\end{equation}
Here, $|\rangle$ denotes the vacuum state in the hole space. 
It is useful to rewrite this expression as the block matrix
\begin{equation}
\label{densmat}
    \hat{\rho_{\mathrm{H}}} = \frac{1}{1+t^2}
    \begin{bmatrix}
    1 & 0 \\
    0 & \hat{\rho}_2 
    \end{bmatrix} \; .
\end{equation}
Here, the two-hole--two-hole matrix $\hat{\rho}_2$ has elements
\begin{equation}
    \rho_{ij}^{kl} = \sum_{a<b} t_{ij}^{ab} t_{kl}^{ab} \ .
\end{equation}
We have $i<j$ and $k<l$ and the matrix $\hat{\rho}_2$ has dimension $D\equiv N(N-1)/2$ 
for a system with $N$ fermions.
As a check, we see that  
\begin{equation}
    \operatorname{Tr}\hat{\rho}_2 = \sum_{i<j}\rho_{ij}^{ij}=t^2 \ ,
\end{equation}
and we indeed have $\operatorname{Tr}\hat{\rho}_H=1$. 
The expression~(\ref{densmat}) is exact and can be used to numerically compute the entropies of the state~(\ref{CCDex}) using Eqs.~(\ref{eq:renyi}) and (\ref{SvN}).

For what follows, we rewrite
\begin{equation}
\label{defsigma}
    \hat{\rho}_2 = t^2 \hat{\sigma} \ , 
\end{equation}
where $\hat{\sigma}$ is a density matrix, i.e. $\operatorname{Tr}\hat{\sigma} = 1$. 

To compute the R{\'e}nyi entropies~(\ref{eq:renyi}) we use 
\begin{align}
    \operatorname{Tr}\hat{\rho}_H^\alpha = (1+t^2)^{-\alpha}\left(1+t^{2\alpha} \operatorname{Tr}{\hat{\sigma}^\alpha}\right) \ .
\end{align}
From here on, we restrict ourselves to $\alpha\ge 1$. We seek further analytical insights and use $t^2\ll 1$. Then,
\begin{equation}
\label{SalCCD}
    S_\alpha = \frac{t^{2\alpha}\operatorname{Tr}\hat{\sigma}^\alpha-\alpha t^2}{1-\alpha} +{\cal O}(t^4) +{\cal O}  (t^{4\alpha}) \ .
\end{equation}
For $\alpha\to 1$ we employ the rule by L'Hospital and find
\begin{equation}
\label{S1CCD}
    S_1 = t^2\left[1 -\operatorname{Tr}\left({\hat{\sigma}\log{\hat{\sigma}}}\right)- \log{t^2} \right] +{\cal O}(t^4) \ .
\end{equation}
The matrix $\hat{\sigma}$ has dimension $D$. Thus, $0\le -\operatorname{Tr}({\hat{\sigma}\log{\hat{\sigma}}})\le \log{D}$. Here, the minimum arises when all but one eigenvalue of $\hat{\sigma}$ vanish, while the maximum arises when all eigenvalues are equal.
Equations~(\ref{SalCCD}) and (\ref{S1CCD}) are the main results of this Section. As we have assumed that $t^2\ll 1$,
\begin{equation}
\label{Sge2}
    S_\alpha = \frac{\alpha}{\alpha-1} t^2 +{\cal O}(t^{2\alpha}) +{\cal O}(t^{4}) \quad\mbox{for $\alpha > 1$,} 
\end{equation}
i.e. the R{\'e}nyi entropies become independent of the eigenvalues of the matrix~(\ref{defsigma}) for sufficiently large index $\alpha$. 

The entropies~(\ref{SalCCD}) and (\ref{S1CCD}) further simplify for arbitrarily weak interactions (i.e. for $t^2\to 0$), and we find the asymptotic behavior
\begin{eqnarray}
\label{asymp}
    S_{\alpha} \to \left\{
    \begin{array}{ll}
    -t^2\log{t^2} & \mbox{for $\alpha=1$ and $t^2\to 0$}\ ,\\
     &\\
    \dfrac{\alpha}{\alpha-1}t^2 & \mbox{for $\alpha>1$ and $t^2\to 0$}\ . 
    \end{array}\right.
\end{eqnarray}
Note that the asymptotic results are independent of the matrix $\hat{\sigma}$ in Eq.~(\ref{defsigma}). The derivation of these results also makes clear that the limits $\alpha\to 1$ and $t^2\to 0$ do not commute.

\subsection{Particle numbers in the hole space}
The number operator for the particles in the hole space is
\begin{equation}
    \hat{N}_\mathrm{H} =\sum_{i=1}^N \hat{a}^\dagger_i\hat{a}_i \ .
\end{equation}
Its matrix representation (limiting the basis to up to two holes) is 
\begin{equation}
\hat{N}_\mathrm{H} = \begin{bmatrix}
    N & 0 \\
    0 & N-2
    \end{bmatrix} \ . 
\end{equation}
This matrix has the same block structure (and dimensions) as $\hat{\rho}_\mathrm{H}$ in Eq.~(\ref{densmat}). Thus, 
\begin{eqnarray}
\label{expNH}
    \langle N_\mathrm{H}\rangle &\equiv& \operatorname{Tr}(\hat{\rho}_\mathrm{H} \hat{N}_\mathrm{H})\nonumber\\
    &=& N-2t^2 +{\cal O}(t^4) \ , 
\end{eqnarray}
and
\begin{eqnarray}
    \langle N_\mathrm{H}^2\rangle &\equiv& \operatorname{Tr}(\hat{\rho}_\mathrm{H} \hat{N}_\mathrm{H}^2)\nonumber\\
    &=&N^2-4t^2(N-1) +{\cal O}(t^4) \ , 
\end{eqnarray}
and the particle-number fluctuation is
\begin{eqnarray}
\label{Nfluc}
    (\Delta N_\mathrm{H})^2 &\equiv& \langle N_\mathrm{H}^2\rangle-\langle N_\mathrm{H}\rangle^2\nonumber\\
    &=& 4t^2 +{\cal O}(t^4) \ .
\end{eqnarray}
Thus, $t^2\approx (\Delta N_\mathrm{H})^2/4$, and substituting this expression into Eqs.~(\ref{SalCCD}) and (\ref{S1CCD}) shows that the R{\'e}nyi entropies [and their asymptotic expressions~(\ref{asymp})] are functions of the particle-number fluctuation. These expressions extend the pioneering results~\cite{klich2006} to finite systems of interacting fermions. 

As it will turn out below, calculations of the expectation value~(\ref{expNH}) are much simpler than computations of the particle-number fluctuation~(\ref{Nfluc}) or the entanglement entropy. In particular, the depletion number of the reference state~\cite{dickhoff2005}
\begin{eqnarray}
\label{depletion}
    \delta N_\mathrm{H} &\equiv& N-\langle N_\mathrm{H}\rangle\nonumber\\
    &=& 2t^2 +{\cal O}(t^4)
\end{eqnarray}
is simple to compute in interacting many-body systems, and this also allows us  to express the entanglement entropy as a function of this quantity.
Thus,
\begin{equation}
\label{t2andNH}
    {1\over 4} (\Delta N_\mathrm{H})^2 \approx  {1\over 2} (\delta N_\mathrm{H}) \approx t^2
\end{equation}
and corrections to this relation are higher powers of $\delta N_\mathrm{H}$  or $(\Delta N_\mathrm{H})^2$ or $t^2$. 

The proportionality between the entropy and the particle-number fluctuation breaks down when one includes higher powers of $T_2$ in the approximation of the CCD ground state~(\ref{CCDex}).  Our analytical results~(\ref{SalCCD}), (\ref{S1CCD}), and (\ref{asymp}), combined with (\ref{t2andNH}) generalize the result~\cite{klich2006} to weakly interacting finite Fermi systems. 

\section{Nuclear systems}
\label{sec:nuclear}
\subsection{Pairing model}
\label{Sec:PM}
The exactly solvable pairing model~\cite{dukelsky2004}
is useful for studying entanglement entropy. The  model consists of $\Omega/2$ doubly degenerate and equally spaced orbitals with two possible spin states $\sigma = \pm 1$. The Hamiltonian is
\begin{equation}
\begin{aligned}
\hat{H} =& \delta \sum_{p \sigma}(p-1) a_{p \sigma}^{\dagger} a_{p \sigma} \\
&-\frac{1}{2} g \sum_{p q} a_{p+}^{\dagger} a_{p-}^{\dagger} a_{q-} a_{q+} \ .
\end{aligned}
\end{equation}
with $p, q = 1, 2,\ldots , \Omega/2$. We set orbital spacing $\delta = 1$ without losing generality, i.e.  all energies (and the coupling $g$) are measured in units of $\delta$. 

We consider the model at half filling with orbitals being either empty or doubly occupied. For sufficiently small coupling strengths, the CCD approximation accurately solves the pairing model~\cite{Lietz2017}. 

\begin{figure}
\centering
\includegraphics[width=0.99\linewidth]{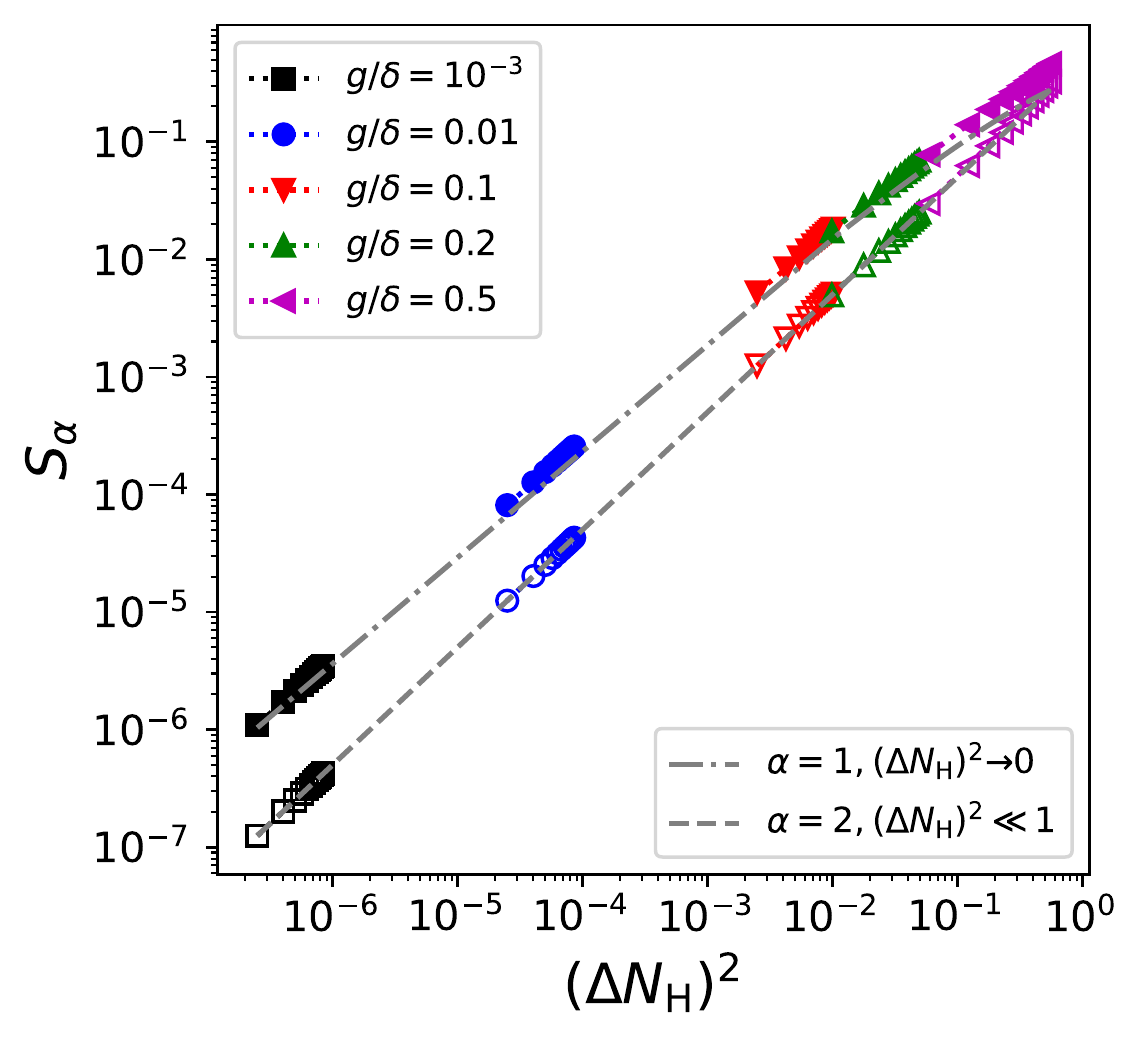}
\caption{R{\'e}nyi entropy $S_1$ (full markers) and $S_2$ (hollow markers) of the reduced hole-space density matrix $\rho_\mathrm{H}$ versus the particle-number fluctuation $(\Delta N_\mathrm{H})^2$ of the hole space for the half-filled pairing model, with $\delta = 1.0$ and different couplings $g$ as indicated. The dash-dotted and dashed lines show analytical results for $\alpha=1$ and $\alpha=2$, respectively, and they are valid for values of $t^2$ as indicated. The color and shape of the markers indicate the coupling strength, and for a given coupling, identical markers show the results for one to twelve pairs. The entropy increases with the number of pairs and with increasing coupling strength.}
\label{fig:PM-Salpha}
\end{figure}

We solve the doubles amplitudes $t_{ij}^{ab}$  using Eq.~(\ref{CCamps}) with $\langle \Phi_{ij}^{ab}|$ as the bra state. We then compute the reduced density matrix~(\ref{densmat}) and the R{\'e}nyi entropy~(\ref{eq:renyi}). For the computation of the von Neumann entropy~(\ref{SvN}) we diagonalize the reduced density matrix. The results are shown in Fig.~\ref{fig:PM-Salpha}. The full and hollow markers are results for $\alpha=1$ and $\alpha=2$, respectively, and the dash-dotted and dashed line are the analytical results~(\ref{Sge2}) and (\ref{asymp}), respectively, combined with Eq.~(\ref{t2andNH}). The different coupling strengths $g$ are identified by the colors and shapes of the markers.  Identical markers show the results of systems containing one to twelve pairs. Entropies (and particle-number fluctuations) increase with coupling strengths and with an increasing number of pairs.  Overall we see that our analytical results agree with data for sufficiently weak interactions, i.e. sufficiently small values of $(\Delta N_\mathrm{H})^2$.

\begin{figure}
\centering
\includegraphics[width=0.99\linewidth]{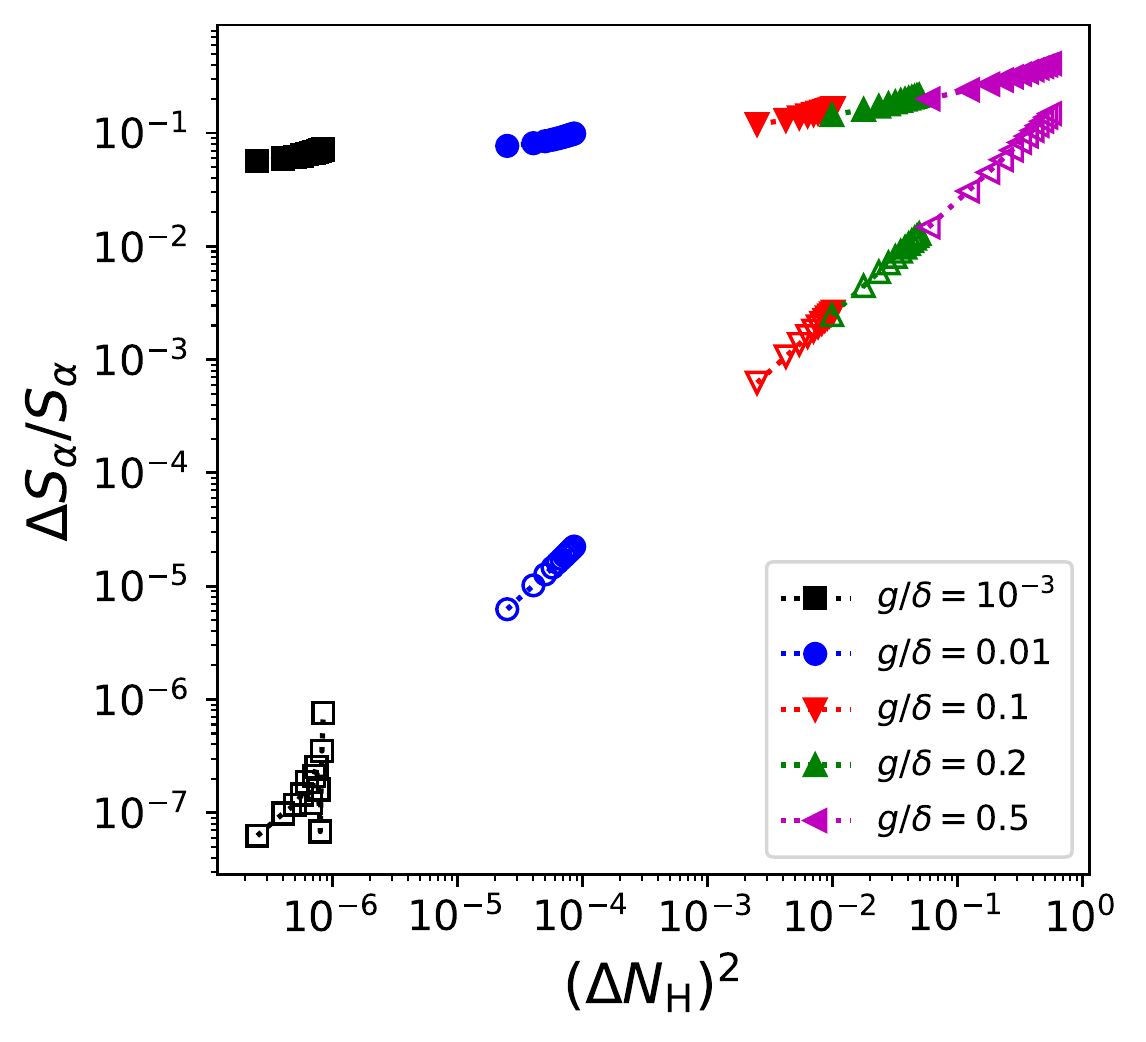}
\caption{Absolute differences between numerical and analytical R{\'e}nyi entropy for $S_1$ (full markers) and $S_2$ (hollow markers), normalized by the numerical entropy,  versus the particle-number fluctuation $(\Delta N_\mathrm{H})^2$ of the hole space for the half-filled pairing model, with $\delta = 1.0$ and different couplings $g$ as indicated. The color and shape of the markers indicate the coupling strength, and for a given coupling, identical markers show the results for one to twelve pairs.}
\label{fig:PM-Salpha-diffs}
\end{figure}

The agreement between numerical and analytical results can be examined closer when plotting the absolute differences between them, normalized by the numerical results. This is shown in Fig.~\ref{fig:PM-Salpha-diffs}. We see that the analytical result for $S_1$ is probably only reached asymptotically for $(\Delta N_{\mathrm{H}})^2\to 0$; this is expected also from Fig.~\ref{fig:PM-Salpha}. We also see that the difference $\Delta S_2$ between the numerical and analytical results is as predicted of order $S_2^2$. We attribute the visible deviations from this behavior for $g/\delta=10^{-3}$ to numerical precision limits, noting that $\Delta S$ is close to machine precision.

A key question is, of course, how the entanglement entropy scales with increasing system size. We can answer that question analytically for small interaction strengths $g/\delta$ by using second-order perturbation theory. We write the cluster amplitudes $t_{i j}^{a b}$ as
\begin{equation}
\label{eq:MBPT2}
    t_{i j}^{a b} \approx \frac{\langle a b|\hat{v}| i j\rangle}{\varepsilon_{i j}^{a b}} \ ,
\end{equation}
where $\varepsilon_{i j}^{a b} = \varepsilon_{i}+\varepsilon_{j}-\varepsilon_{a}-\varepsilon_{b}$ and $\varepsilon_p \equiv (p-1)\delta$ for the pairing model. Thus, 
\begin{equation}
\begin{aligned}
\label{eq:t2_A}
    t^2 & = \frac{1}{4}\sum_{i=1}^{\frac{N}{2}} \sum_{a=\frac{N}{2}+1}^{\frac{\Omega}{2}} \frac{g^2}{4\delta^2(i-a)^2} \\
    &\approx \frac{g^2}{16\delta^2}\sum_{i=1}^{\frac{N}{2}} \left[ \int_{\frac{N}{2}+1}^{\frac{\Omega}{2}} \frac{1}{(i-a)^2}da \right] \\
    &\approx \frac{g^2}{16\delta^2} \int_1^{\frac{N}{2}} \left[\frac{1}{i-\frac{\Omega}{2}} - \frac{1}{i-\frac{N}{2}-1}\right]di\\
    & = \frac{g^2}{16\delta^2} \log{\frac{N(\Omega-N)}{2(\Omega-2)}}\\
    &\approx \frac{g^2}{16\delta^2} \log{\frac{N}{4}}
\end{aligned}
\end{equation}
where $N = \Omega/2$ at half filling. Here the last step is valid when $N\gg 1$, and we approximated the sums  by integrals using the Euler–Maclaurin formula. This approximation introduces an error of order ${\cal O}(N^0)$. 

\begin{figure}
\centering
\includegraphics[scale=0.6]{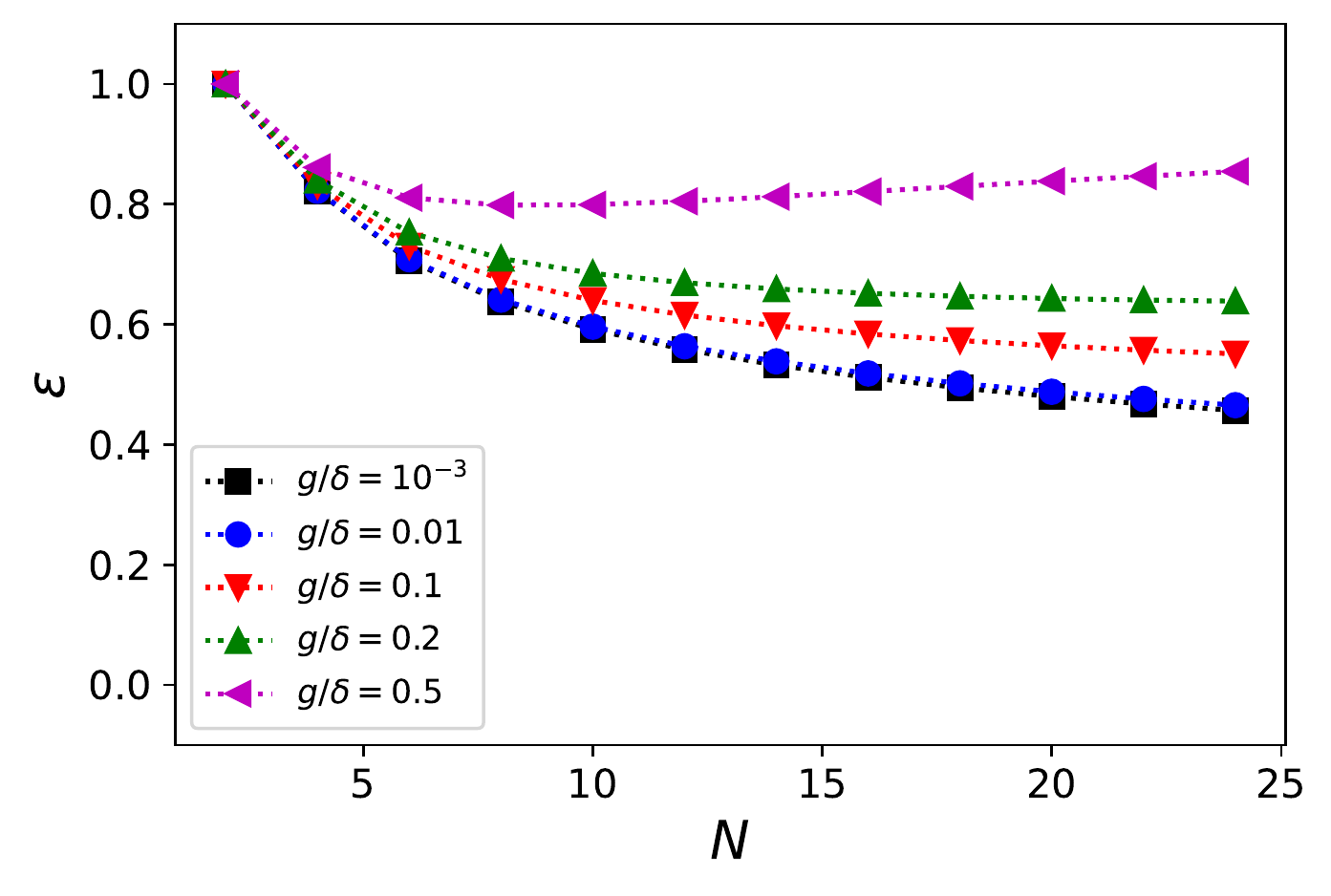}
\caption{Error of the approximation over number of particles, with $\delta = 1.0$ and $g = 1e-4, 1e-3, 1e-2, 1e-1, 2e-1, 5e-1$.}
\label{fig:t2vsg}
\end{figure}

To see this, we compute the relative error at half filling ($\Omega=2N$)
\begin{equation}
    \varepsilon = \frac{\left|t^2 - \frac{g^2}{16\delta^2}\log{N^2\over 4(N-1)}\right|}{t^2} \ , 
\end{equation}
and show the result in Fig.~\ref{fig:t2vsg}.
We can see that for small enough $g$, Eq.~\eqref{eq:MBPT2} is valid, and $t^2 \propto \log(N)$ is the leading approximation. Thus for $\alpha \geq 2$ we have $S_\alpha \propto \log(N)$. This agrees with expectations for a Fermi system in one dimension~\cite{leschke2014}.

\subsection{Neutron matter}
\label{Sec:NM}
Neutron matter is relevant to understand neutron-rich nuclei and neutron stars. Here, we consider a simple yet non-trivial model of  neutron matter based on the the  Minnesota potential~\cite{Thompson1977}. This is a simplification from more realistic descriptions, e.g. within chiral effective field theory, and only employs two-body forces. The Hamiltonian consists of the kinetic energy $\hat{t}_0$ and the Minnesota potential $\hat{v}$ 
\begin{equation}
    \hat{H}=\hat{H}_{0}+\hat{H}_{I}=\sum_{i=1}^{A} \hat{t}_{0}\left(x_{i}\right)+\sum_{i<j}^{A} \hat{v}\left(r_{i j}\right) \;.
\end{equation}
The  Minnesota potential consists of a repulsive core and a short-range attraction employing the exponential functions   $\operatorname{exp}(-\alpha_i r^2)$ of the two-particle distance $r$.  
We compute neutron matter using a basis consisting of discrete momentum states $|k_x,k_y,k_z\rangle$ in a cubic box with periodic boundary conditions. This follows the coupled-cluster calculations of Ref.~\cite{Lietz2017}, with the Python notebook~\cite{papenbrock2018}. 

The number of cubic momentum states is $(2 N_\mathrm{max}+1)^3$. The spin degeneracy for each momentum state is $g_{\rm st} = 2$. We limit our calculation to neutron matter with density $n \approx 0.08$ $\mathrm{fm}^{-3}$; this is about half of the saturation density of nuclear matter. Using $N$ neutrons, the volume is $L^3$ with $L = (N/n)^{1/3}$, and we employ closed-shell configurations of $N = 14,38,54,66,114$ particles in our calculation. Details about the basis space are presented in Refs.~\cite{hagen2013b,papenbrock2018}.

We use a simplified version of the coupled-cluster with doubles approximation based on ladder diagrams only. This is sufficiently accurate for the Minnesota potential~\cite{hagen2013b} and agrees with virtually exact results from the auxiliary field diffusion Monte Carlo (AFDMC) method~\cite{gandolfi2009}. 

The relevant matrix elements of the similarity transformed Hamiltonian $e^{-T_2}He^{T_2}$ are 
\begin{equation}
\begin{aligned}
    \bar{H}_{ij}^{ab} &= \left\langle \vec{k_a}\vec{k_b}|v|\vec{k_i}\vec{k_j}\right\rangle \\
    &+ P(ab)\sum_c f_c^b t_{ij}^{ac} \\
    &- P(ij)\sum_k f_j^k t_{ik}^{ab}\\
    &+ \frac{1}{2}\sum_{cd} \left\langle \vec{k_a}\vec{k_b}|v|\vec{k_c}\vec{k_d}\right\rangle  t_{ij}^{cd} \\
    &+ \frac{1}{2}\sum_{kl} \left\langle \vec{k_k}\vec{k_l}|v|\vec{k_i}\vec{k_j}\right\rangle  t_{kl}^{ab} \ .
\end{aligned}
\end{equation}
Here we introduced the Fock matrix with elements
\begin{equation}
    f^p_q = \left\langle \vec{k_p}| t_0 |\vec{k_q}\right\rangle + \sum_i \left\langle \vec{k_p}\vec{k_i}| v |\vec{k_q}\vec{k_i}\right\rangle \ , 
\end{equation}
and $P(pq)$ is a permutation operator. Solving the equation $\bar{H}_{ij}^{ab}=0$ yields the amplitudes $t_{ij}^{ab}$.

\begin{figure}
\centering
\includegraphics[scale=0.6]{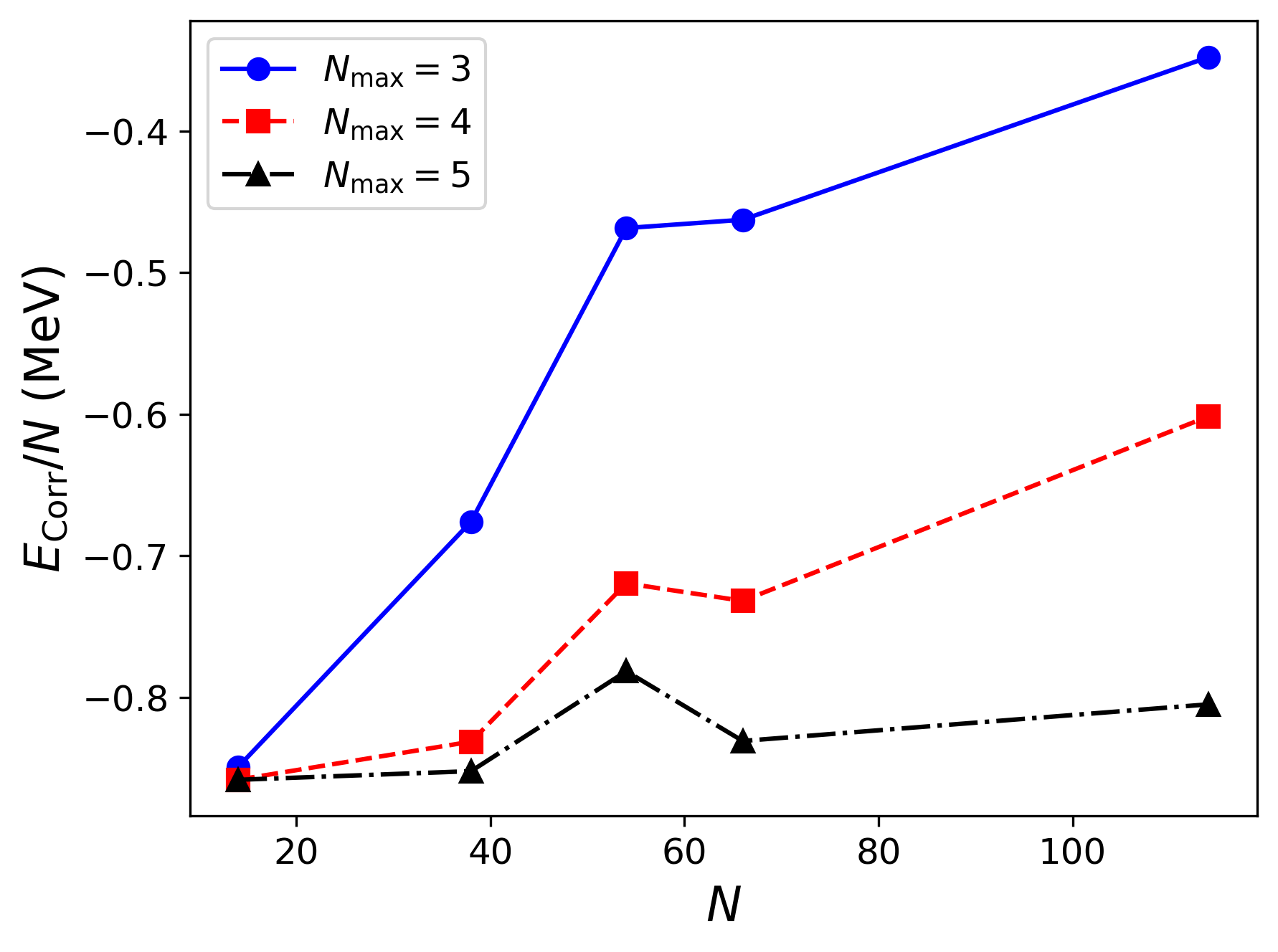}
\caption{Correlation energy per neutron versus the neutron number $N = 14, 38, 54, 66, 114$ with different size $N_{\text{max}}$ of momentum space.}
\label{fig:NM_CorrEperN_A}
\end{figure}

Figure~\ref{fig:NM_CorrEperN_A} shows the correlation energy per neutron as a function of neutron number. The correlation energy is defined as the difference between the CCD energy~(\ref{Eccd}) and the  Hartree-Fock energy $E_{\text{HF}}$ 
\begin{equation}
    E_{\text{HF}} = \sum_{i} \left\langle\vec{k_i}  |t_0|\vec{k_i}\right\rangle + \frac{1}{2}\sum_{i,j} \left\langle\vec{k_i}\vec{k_j}  |v|\vec{k_i}\vec{k_j}\right\rangle
\end{equation}
of the reference state. We see that the correlation energy depends weakly on $N$ (and becomes approximately constant) for $N_{\mathrm{max}}=5$. We attribute the peak at $N=54$ to finite-size effects, i.e. shell oscillations. We note that these shell oscillations can be reduced using twist-averaged boundary conditions~\cite{gros1996,lin2001,hagen2013b}.   

Table~\ref{tab:tsq_NM} shows the value of $t^2$ from Eq.~(\ref{eq:t2}) for various $N_\mathrm{max}$. We see that $t^2 \ll 1$, required for the applicability of our analytical results regarding entropies, is only valid for for $N \lesssim 66$.  Thus, we limit the analysis to $N\leq66$ for neutron matter. 
\begin{table}[ht]
\begin{tabular}{l|r|r|r|r|r}
 & $N=14$ &  $N=38$ &  $N=54$ &  $N=66$&  $N=114$ \\ \hline
$N_\mathrm{max}=3$ & 0.106 & 0.298 & 0.246 & 0.475 & 1.239 \\ \hline
$N_\mathrm{max}=4$ & 0.106 & 0.322 & 0.299 & 0.557 & 1.431\\
\hline
$N_\mathrm{max}=5$ & 0.106 & 0.324 & 0.308 & 0.581 & 1.565\\ \hline
\end{tabular}
\caption{Numerical values for $t^2$ for different neutron matter models $N = 14, 38, 54, 66, 114$ with increasing momentum space size.}
\label{tab:tsq_NM}
\end{table}

\begin{figure}
\centering
\includegraphics[scale=0.6]{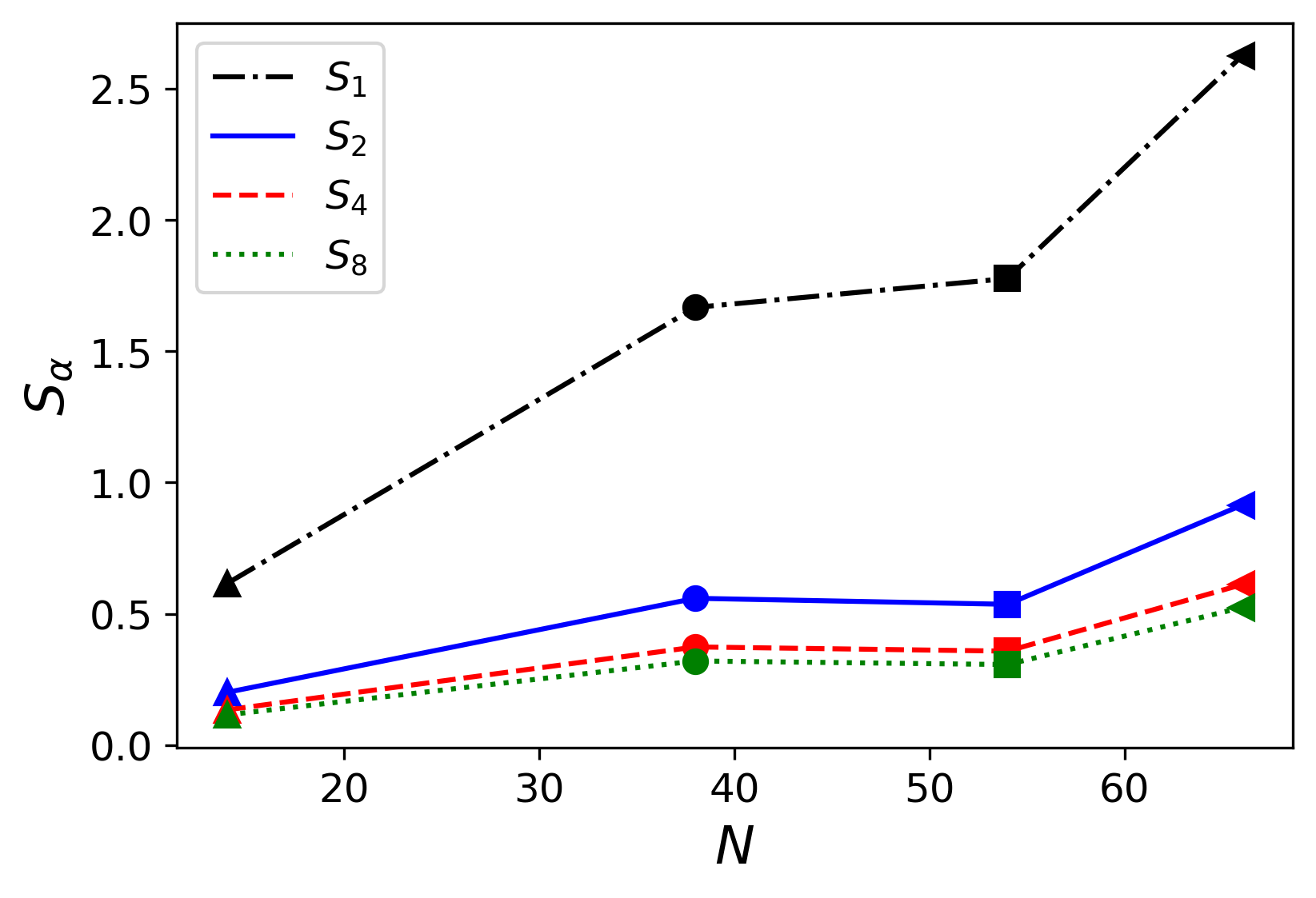}
\caption{R\'enyi entropy (von~Neumann entropy $S_1$ is denoted as limiting case of R\'enyi entropy) versus the neutron number $N = 14$ (triangle\_up), $N = 38$ (circle), $N = 54$ (square), $N = 66$ (triangle\_left), $N_{\text{max}} = 5$ of momentum space.}
\label{fig:NM_entropy_A}
\end{figure}

We compute the entanglement entropies by partitioning the single-particle basis as follows: The Fermi sphere, i.e. the set of lattice sites occupied in the Hartee-Fock state of a  closed-shell configuration, is the hole space, and all other lattice sites are the particle space. 
Figure~\ref{fig:NM_entropy_A} shows R{\'e}nyi entanglement entropies $S_\alpha$ for $\alpha=1,2,4$ and $8$ of neutron matter  as a function of the neutron number $N$. The entropies increase approximately linearly with increasing neutron number (and $N=54$ is again an outlier). This is expected because the short-range Minnesota potential couples the Fermi sphere to all momentum states in the particle space. Thus, a volume law holds for neutron matter in momentum space. 

\begin{figure}
\centering
\includegraphics[scale=0.6]{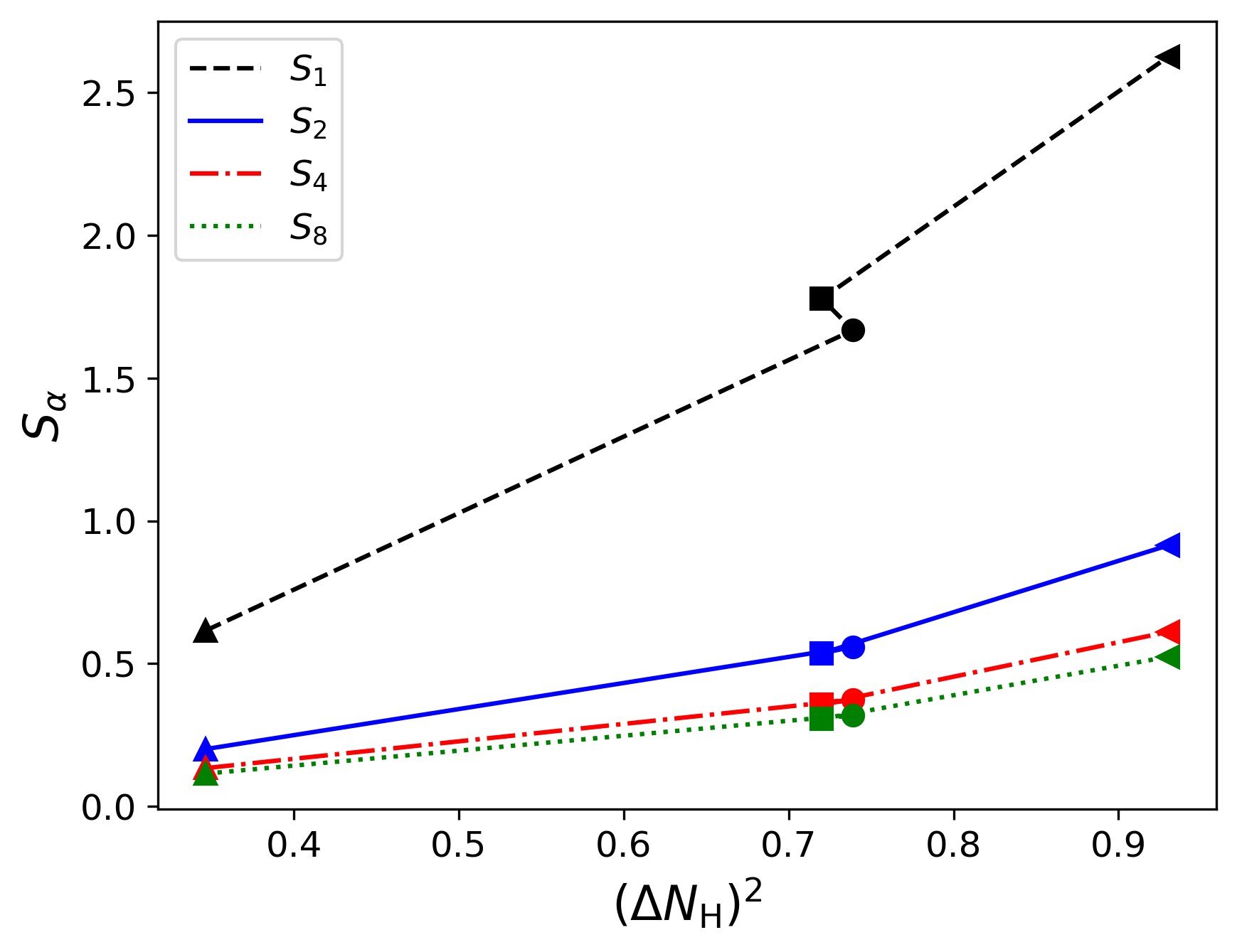}
\caption{R\'enyi entropy (von~Neumann entropy $S_1$ is denoted as limiting case of R\'enyi entropy) versus the particle number variation with $N = 14$ (triangle up), $N = 38$ (circle), $N = 54$ (square), $N = 66$ (triangle left), $N_{\text{max}} = 5$ of momentum space.}
\label{fig:NM_entropy_Var}
\end{figure}

Figure~\ref{fig:NM_entropy_Var} shows the entanglement entropies versus the particle number fluctuations. Again, the relation is approximately linear.  

The results of this Section show that neutron matter exhibits entanglement entropies (in momentum space) that are approximately proportional to the neutron number; they are also approximately proportional to the particle-number fluctuations. The latter result is less accurate than for the pairing model. This is because the size of the $T_2$ amplitudes is sizeable. i.e. we have $t^2< 1$ but not really $t^2\ll 1$.      

\subsection{Finite nuclei}
Computing the entanglement entropy in finite nuclei is a computationally daunting task: model spaces consist of ${\cal O}(1000)$ of single-particle states, and the hole-space density matrix required for this task is a many-body operator. Instead, we use the depletion number~(\ref{depletion}) as an entanglement witness, because for small cluster amplitudes, the depletion number is proportional to the R{\'e}nyi entropies, see Eqs.~(\ref{Sge2}) and (\ref{t2andNH}). The depletion number can be accurately computed with coupled-cluster theory, as we describe in the following paragraph. In contrast, the particle-number fluctuation of the hole space is a small number resulting from cancellations of two large numbers. Being non-Hermitian, the coupled-cluster method does not guarantee that the particle-number variation is non-negative. 

We perform coupled-cluster singles-and-doubles (CCSD) computations of the closed-shell nuclei $^4$He, $^{16}$O, $^{40}$Ca, and $^{100}$Sn using the interactions of Ref.~\cite{hebeler2011}. The CCSD approximation accounts for about 90\% of the correlation energy and is a size-extensive method, i.e. the error in the correlation energy is proportional to the mass number $A$. For the calculations, we employ a model space of 15 major harmonic oscillator shells and use an oscillator spacing of $\hbar\omega=16$~MeV. We perform a Hartree-Fock computation to obtain the reference state $|\Phi\rangle$, and this defines the hole space. We then solve the CCSD equations, and compute the similarity-transformed Hamiltonian $\overline{H}$ where 
\begin{equation}
\overline{O} \equiv e^{-\hat{T}} \hat{O} e^{\hat{T}}    
\end{equation}
for any operator $\hat{O}$. We solve for the left ground state $\langle L|\equiv \langle\Phi|(1+\hat{\Lambda})$ of $\overline{H}$; here $\hat{\Lambda}$ is a 1p-1h and 2p-2h de-excitation operator. We then compute the hole-space occupation as 
\begin{equation}
    \langle N_\mathrm{H} \rangle = \langle L|\overline{N} |\Phi\rangle \ , 
\end{equation}
and the depletion number becomes
\begin{equation}
\label{Adeplete}
    \delta A = A-\langle N_\mathrm{H} \rangle 
\end{equation}
for a nucleus with the mass number $A$. This approach is valid also for large coupled-cluster amplitudes. 

\begin{figure}
\centering
\includegraphics[width=0.99\linewidth]{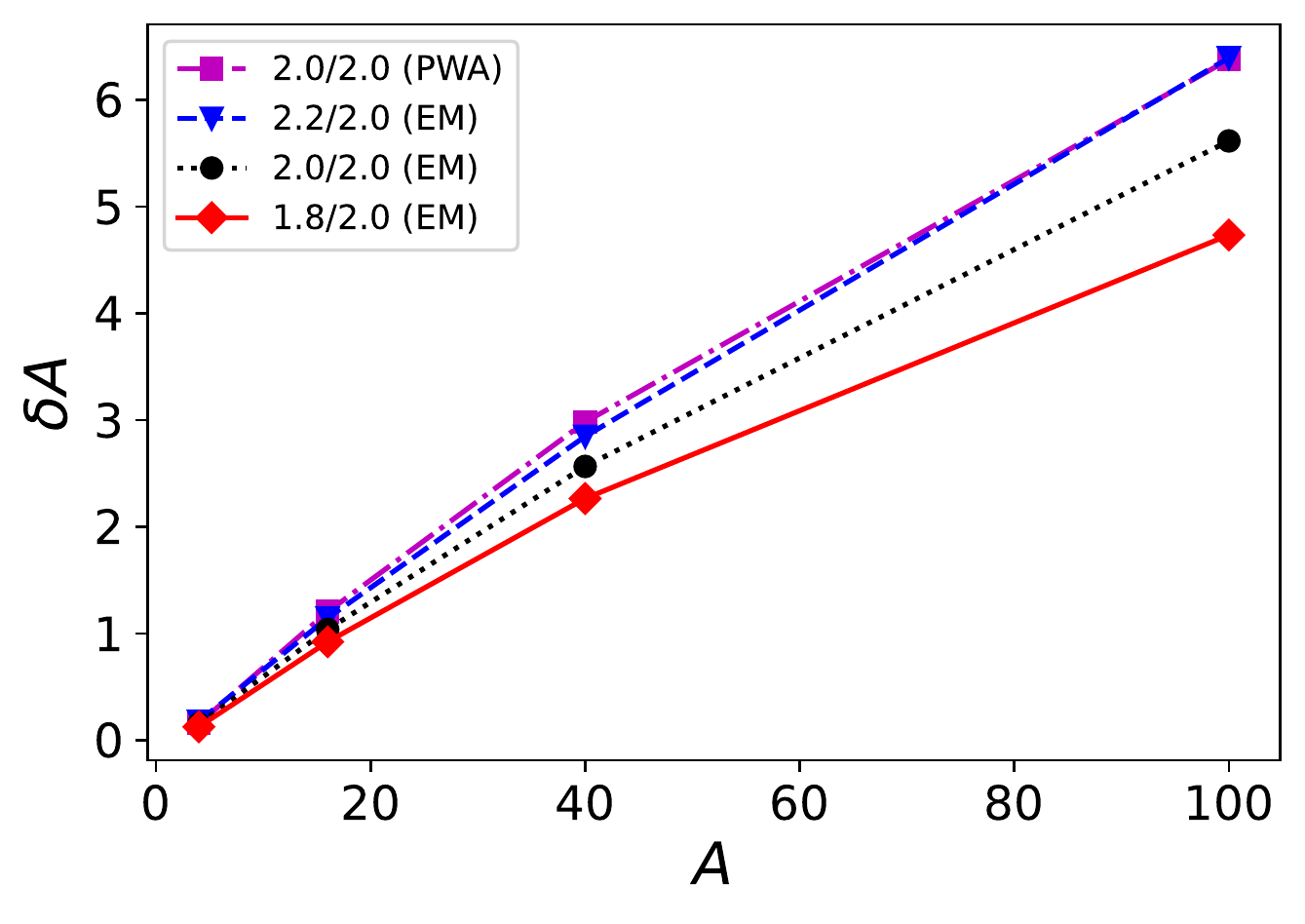}
\caption{Depletion number $\delta A$ of the hole space in the nuclei $^4$He, $^{16}$O, $^{40}$Ca, and $^{100}$Sn computed with the interactions of Ref.~\cite{hebeler2011} as indicated, as a function of the mass number $A$. }
\label{fig:nuclei-depletion}
\end{figure}

Figure~\ref{fig:nuclei-depletion} shows the results for the depletion number~(\ref{Adeplete}) for $^4$He, $^{16}$O, $^{40}$Ca, and $^{100}$Sn computed with the interactions from Ref.~\cite{hebeler2011} as a function of the mass number $A$. The numbers in the labels indicate the values of the momentum cutoffs (in fm$^{-1}$) employed for the two- and three-body interactions, respectively. The depletion number is larger for ``harder'' interactions, i.e. for those with larger momentum cutoffs, and this meets our expectations. We see also that the depletion number approximately is an extensive quantity (i.e. linear in $A$). Its scaling with $A$ is certainly closer to $A^1$ than to $A^{2/3}$, thus preferring a volume over an area law. This is consistent with the arguments presented in Sect.~\ref{sec:intro}. 

\section{Summary}
We studied entanglement in nuclear systems, based on a partition of the single-particle space into holes and particles. This is the most natural choice for finite systems. Analytical arguments based on coupled-cluster theory show that the R{\'e}nyi entropies $S_\alpha$ for $\alpha>1$ are proportional to the number variation and the depletion number of the hole space. This extends analytical arguments for non-interacting fermions to systems with sufficiently weak interactions. For arbitrary weak interactions, we also obtain universal results for the von Neumann entropy $S_1$. 

We confirmed our analytical results using numerical solutions of the pairing model. For a semi-realistic model of neutron matter, we showed that entanglement entropies of the Fermi sphere are approximately proportional to the particle number fluctuations of the hole space and to the number of neutrons. The former  confirms our analytical results and the latter agrees with expectations for short-ranged interactions. 
Finally, we computed the depletion number in finite nuclei using interactions from chiral effective field theory. We saw that the entanglement witness increases with an increasing cutoff of the employed interaction and again grows approximately linear with the mass number.  
\begin{acknowledgments}
This material is based upon work supported by the
U.S.\ Department of Energy, Office of Science, Office of Nuclear Physics under award numbers DE-FG02-96ER40963, DE-SC0021642, and DE-SC0018223 (NUCLEI
SciDAC-4 collaboration), the NUCLEI SciDAC-5 collaboration, and by the Quantum Science Center, 
a National Quantum Information Science Research Center of the U.S. Department of Energy. Computer time was provided by the
Innovative and Novel Computational Impact on Theory and Experiment (INCITE) program. This research used resources from the Oak Ridge Leadership Computing Facility located at Oak Ridge National Laboratory, which is supported by the Office of Science of the Department of Energy under contract No. DE-AC05-00OR22725.
\end{acknowledgments}


%

\end{document}